\documentclass{eas}
\usepackage{graphicx}
\begin{document}

\title{Submillimetre/TeraHertz Astronomy at \\
Dome C with CEA filled bolometer array}\thanks{This work has been undertaken within the
framework of the EC ARENA network activities.}
\author{Vincent Minier}\address{Service d'Astrophysique/DAPNIA/DSM, CEA Saclay, 91191 Gif-sur-Yvette, France}
\author{Gilles Durand}\sameaddress{1}
\author{Pierre-Olivier Lagage}\sameaddress{1}
\author{Michel Talvard}\sameaddress{1}
\author{Tony Travouillon}\address{TMT project, California Institute of Technology, Pasadena 91125 CA, USA}
\author{Maurizio Busso}\address{Dipartimento di Fisica, Universit\`a di Perugia, Via A. Pascoli, I-06100 Perugia, Italy}
\author{Gino Tosti}\sameaddress{3}
\begin{abstract}
Submillimetre/TeraHertz (e.g. 200, 350, 450 $\mu$m) astronomy is the prime technique to unveil the birth and early evolution of a broad range of astrophysical objects. A major obstacle to carry out submm observations from ground is the atmosphere. Preliminary site 
testing and atmospheric transmission models tend to demonstrate that Dome C could offer the best conditions on Earth for submm/THz astronomy. The CAMISTIC project aims to install a filled bolometer-array camera with 16x16 pixels on IRAIT at Dome C and explore the 200-$\mu$m  windows for potential ground-based observations. 
\end{abstract}
\runningtitle{Minier et al.: Submm/THz astronomy at Dome C}
\maketitle
\section{Submm astronomy: science drivers and novel bolometer technology}

Submillimetre (submm) astronomy is the prime technique to unveil the birth and early evolution of a broad range of astrophysical objects. It is a relatively new branch of observational astrophysics which focuses on studies of the Ôcold UniverseÕ, i.e., objects radiating a significant $-$ if not dominant $-$ fraction of their energy at wavelengths ranging from $\sim100$ $\mu$m to $\sim1$ mm.  Submm continuum observations are particularly powerful to measure the luminosities, temperatures and masses of cold dust emitting objects . Examples of such objects include star-forming clouds in our Galaxy, prestellar cores and deeply embedded protostars, protoplanetary disks around young stars, as well as nearby starburst galaxies and dust-enshrouded high-redshift galaxies in the early Universe (Fig. 1). 

\begin{figure}
\centering
\includegraphics[scale=0.06]{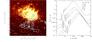}
\caption{Left: NGC7538, a high-mass star-forming region at 2.6 kpc. The grey contours are the 1.2 mm
continuum emission that was imaged with MAMBO-2/IRAM-30m and overlaid on the DSS optical image of the HII region. High-mass protostellar objects are localed in the mm emission clumps to the south of the HII region (Pestalozzi et al. 2006). As high-mass protostellar objects emit the bulk of their energy between 60 and 400 
$\mu$m (Minier et al. 2005), submm/THz continuum mapping is a unique tool to study them. Right: Spectral Energy Density of ultra-luminous galaxies for redshift z=0.1 to 5. Submm/THz continuum observations measure SED peaks for z=3 to 5 (Guiderdoni et al. 1998).}
\end{figure}

Observations of submm continuum emission are usually carried out with bolometer detectors. Recently, CEA (DSM/DAPNIA/SAp and DRT/LETI) has developped a novel bolometer technology for the PACS submm/far-infrared imager on the Herschel Space Observatory. The R\&D was based 
on a unique and innovating technology that combines all silicon technology (resistive thermometers, absorbing grids, multiplexing) and monolithic fabrication. The bolometers are assembled on a mosaic "CCD-like" array that provides full sampling of the focal plane with $\sim2000$ pixels that are arranged in units of 256 pixels. They are cooled down to 300 mK to optimise the sensitivity down to the physical limit imposed by the photon background noise. The PACS bolometer arrays have passed all the qualification tests (Billot et al. 2006). The newly started ArT\'eMiS project at CEA Saclay capitalises on this achievement by developing submm (200-450 $\mu$m) bolometer arrays with $\sim4100$ pixels for ground-based telescopes. A prototype camera operating in the 450 $\mu$m atmospheric window has successfully been tested in March 2006 on the KOSMA telescope (Talvard et al. 2006). 

Placed on a 12-m single-dish telescope that is located on a dry site (25$\%$ PWV$<$0.5 mm) , a bolometer camera with $\sim$4100 pixels at 
200-450 $\mu$m will be particularly powerful to undertake wide field surveys of star-forming complexes in our Galaxy as well as deep field surveys 
of dust-enshrouded high-redshift galaxies in the early Universe (see Talvard et al. 2006 for an overview of ArTeMiS science cases). Beside Herschel and ALMA, large-format submm imagers will provide (1) better angular resolution than Herschel around $\sim200$ $\mu$m and (2) wider-field mapping capabilities than ALMA, making large-scale surveys possible. 

\section{Atmosphere transmission: Chajnantor vs. Dome C}

A major obstacle to carry out submm observations from ground is the atmosphere. Astronomical observations in the submm/THz spectral bands (e.g. 200, 350, 450 $\mu$m) can only be achieved from extremely cold, dry and stable sites (e.g. high altitude plateau, Antarctica) or from space (e.g. the Herschel Space Observatory) to overcome the atmosphere opacity and instability that are mainly due to water vapour absorption and fluctuations in the low atmosphere. Chile currently offers the best accessible (all-year long) sites on Earth, where the precipitable water vapour (PWV) content is often less than 1 mm. Chile hosts the best astronomical facilities such as ESO VLT, APEX and the Chajnantor plateau will be the ALMA site. 
At longer term, and particularly if global warming severely restricts the 200-350-450-$\mu$m windows on ESO sites, Antarctica conditions with less than 0.2 mm PWV, could offer an exciting alternative for THz/submm astronomy (Fig. 2). This is an attractive opportunity for the 
200-$\mu$m windows, especially, which are normally explored with Space telescopes (e.g. Herschel). 

\begin{figure}
\includegraphics[scale=0.45]{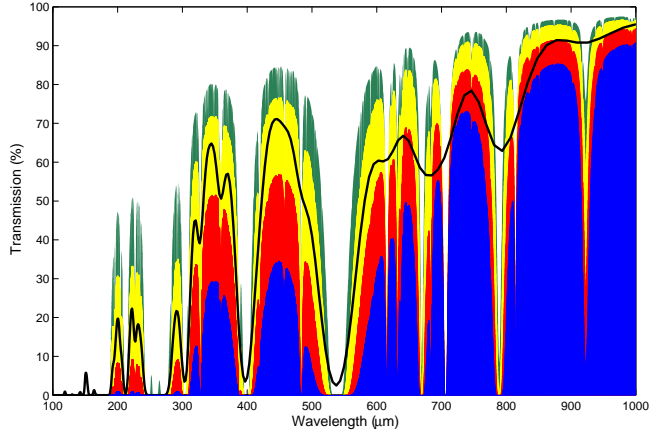}
\caption{Modeled atmospheric transmission for submm/mm wavelengths. Colour plots are ATM models (Pardo
et al. 2001)
for Chajnantor with PWV equal to 0.1 mm (green), 0.2 mm (yellow), 0.5 mm (red) and 1 mm (blue). The solid line represents the atmospheric transmission expected at Dome C for PWV=0.17 mm (Lawrence 2004).}
\end{figure}

In order to determine the respective performances of a submm bolometer array at Chajnantor and Dome C, we have modeled the
atmospheric transmisssion of the two sites. Using the freely available Atmospheric Transmission Model (ATM;
Pardo et al. 2001), we
have computed the transmission statistics at 200 and 450 $\mu$m. As input to the model, PWV of the two sites are necessary. 
In the case of Chajnantor, the PWV statistics available in Peterson et al.
(2003) were used. These statistics span a total period of $\sim5$ years. In the case of Dome C, no PWV data is
available for a long period of time including winter. The closest site where a large amount of statistics are available is the South Pole. 
The difference between the South Pole and Dome C, in terms of PWV are expected to be small. Transmission measurements in the summer time at 350
$\mu$m made at both sites by Calisse et al. (2004) suggest that the transmissions are similar while Dome C has better
stability than the South Pole. For this model, we therefore use the PWV data from the South Pole available
in Peterson et al. (2003) and extrapolate the corresponding atmospheric transmission at Dome C using the model in Lawrence et al. (2004). Our transmission calculations for Dome C must therefore be considered conservative.

Fig. 3-left shows the results of the transmission statistics at Dome C and Chajnantor for 450 $\mu$m. At this
wavelength, the transmission of the two sites is only comparable for the best 10 percent of the time. 
For non-exclusive use of telescope time for this wavelength, Chajnantor is therefore the optimal site 
as sub-mm observatories (e.g. APEX, ASTE, NANTEN-2) are already in place and is therefore cheaper than 
Dome C were no sub-mm facilities are currently available. For exclusive use of telescope time for the 450 $\mu$m
window, Dome C becomes however a more attractive site. This window remains above a transmission of 50\% for more 
than 90\% of the time while the transmission at Chajnantor drops far quicker to a median
value of 35\%. At 200 $\mu$m, this difference is even more severe. As shown in Fig. 3-right, observing from 
Chajnantor requires exceptional conditions on the APEX site at this wavelength. Even at Dome C, the best conditions 
are required to attempt the observation of bright sources. 

\begin{figure}
\centering
\includegraphics[scale=0.41]{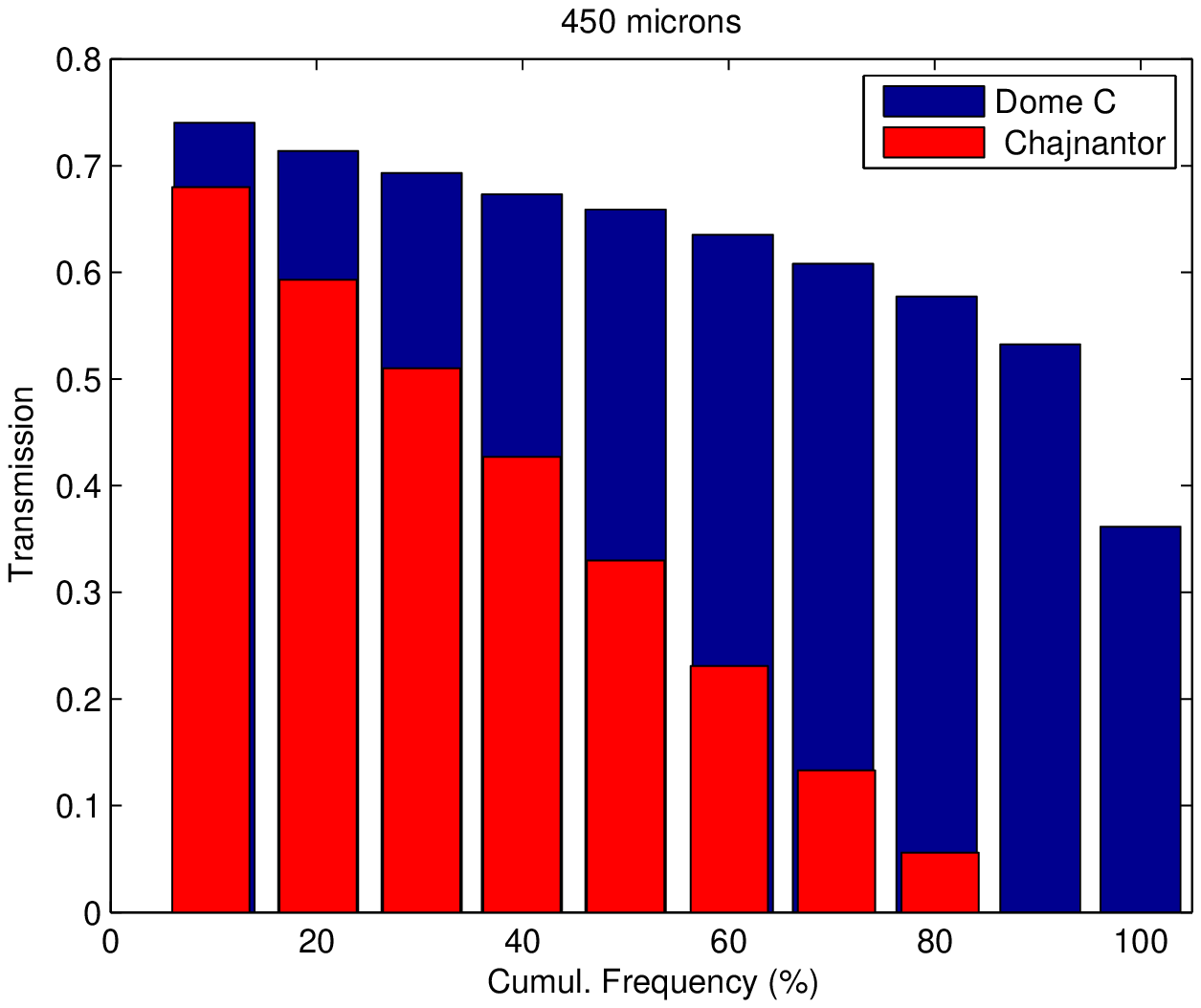}
\includegraphics[scale=0.41]{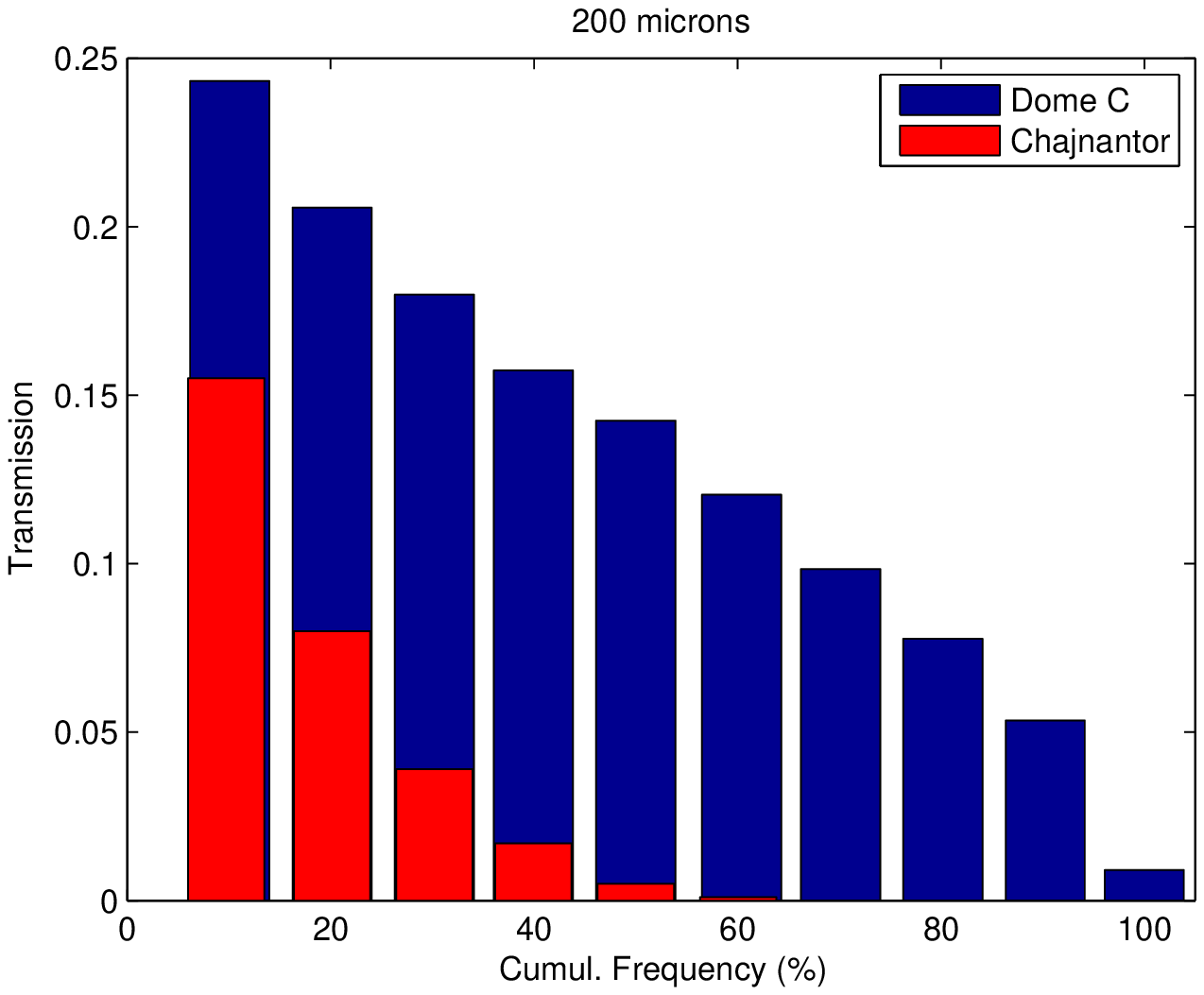}
\caption{Atmospheric transmission vs. cumulative time frequency for Chajnantor (red) and Dome C (blue). Left: 450 $\mu$m. Dome C modeled
transmission remains above 50\% during 90\% of the time. Right: 200 $\mu$m. The window opens only less than 10\% of the time for a transmission
above 20\%.}
\end{figure}

The stability of these windows is an equally important parameter when comparing the sites. Fast variations of the
PWV mean that the transmission can vary significantly during an observation. Peterson et al. (2003) have calculated the
statistics of the standard deviation of the PWV over periods of 2 hours (typical observation time) for South Pole and Chajnantor.
Using the 25\% and 50\% percentile of the distribution of this number for the two sites, we have calculated the corresponding
standard deviation of the transmission at 200 and 450 $\mu$m as a definition for stability and applied the South Pole values
to Dome C modeled transmission in Fig. 3. These results are
shown in Fig. 4. In this figure, the error bars correspond to the standard deviation of the transmission. The model for Dome C 
stands out as being far more stable than Chajnantor
where the transmission in median stability conditions can varies by up to 20\%. It is only at 200 $\mu$m, for the
worse 50\% stability conditions that the stability at Dome C becomes an issue.

The Antarctic plateau is therefore a potentially ideal site for submm/THz observations. The conditions are probably always
ideal for observations at 450 $\mu$m and is a potential ground based site where 200 $\mu$m observations
can be made paving the path for complementary observations with higher angular resolution to the Herschel Space mission.

\begin{figure}
\centering
\includegraphics[scale=0.45]{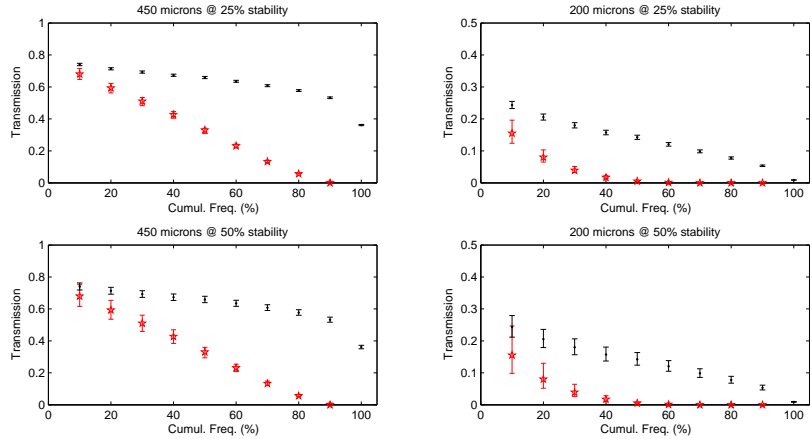}
\caption{Atmospheric transmission vs. cumulative frequency at Dome C (black circles) and Chajnantor (red stars). 
The error bars correspond to the modeled standard deviation of the transmission for the  25\% and 50\% stability percentiles.
25\% stability means the stability level that is expected 25\% of the winter time.}
\end{figure}

\section{Site testing and science qualification}

Chajnantor plateau in Chile is currently presented as the best submm astronomy site on Earth. However, our analysis in Sect. 2 predicts that 
the French-Italian Concordia base could be a potentially remarkable observatory on Earth for submm/THz astronomy.  A site testing campaign that is specifically designed for submm/THz astronomy is clearly required. The transparency (i.e. the optical depth $\tau$) and the stability (i.e. the sky noise ${\sigma}_{\tau}$) must be measured to determine whether Dome C is a better site than Chajnantor. 

The road map for site testing that we propose to follow at CEA Saclay will consist of three steps. (1) The GIVRE
experiment (2007) in collaboration with IPEV and LUAN, Nice aims to determine the impact of frost and icing on hardware at the Concordia station. (2) The SUMMIT 200 $\mu$m tipper experiment (2008) in collaboration with UNSW, Sydney aims to  measure ${\tau}$ and therefore obtain a clear estimate of the atmosphere transmission. (3) The CAMISTIC project (2009-2010) at CEA Saclay aims to install a filled bolometer-array camera with 16x16 pixels on the IRAIT telescope (Tosti et al. 2006) at Dome C  and explore the 200-$\mu$m (i.e. THz) windows for ground-based observations. CAMISTIC will perform site testing on the atmospheric transmission and sky noise in the 200-$\mu$m range.  Sky noise measurements will be performed through sky imaging with the whole bolometer array. On a
80-cm telescope (IRAIT), the near field is located at 1.6 km for 200 $\mu$m. This means that the overlap between the bolometer pixel beams will be greater than 50\% until about 1 km, and above the near-field limit, the pixel beams completely diverge. If the water vapour cells responsible for the fluctuations in the atmosphere background power (i.e. sky noise) are located below $\sim1$ km, the pixels "see" through the same atmosphere column and sky noise. However, if these water vapour cells are above 1.6 km, then non adjacent pixels on the array "see" through different atmosphere column and can potentially provide a map of the sky noise. 

In the future this site testing campaign should demonstrate whether Dome C is the best site on Earth for submm/THz astronomy and therefore open the
path to large ($>10$) diameter telescopes at the Concordia station (see Olmi et al., this volume).


\end{document}